\documentclass[aps,twocolumn,showpacs,preprintnumbers,amsmath,amssymb]{revtex4-1}
\usepackage{latexsym}
\usepackage{graphicx}
\usepackage{dcolumn}
\usepackage{bm}
\usepackage{xcolor}
\usepackage{float}
\usepackage{enumitem}
\usepackage{amssymb,amsmath}

\usepackage{CJK} 

\newcommand{\half}{{{\textstyle\frac{1}{2}}}}
\newcommand{\quarter}{{{\textstyle\frac{1}{4}}}}
\newcommand{\be}{\begin{equation}}
\newcommand{\ee}{\end{equation} }
\newcommand{\beqa}{\begin{eqnarray} }
\newcommand{\eeqa}{\end{eqnarray} }
\newcommand{\ba}{\begin{array}}
\newcommand{\ea}{\end{array}}

\newcommand{\so}{\mathbf{so}}
\newcommand{\SO}{\mathbf{SO}}
\newcommand{\Spin}{\mathbf{Spin}}
\newcommand{\GL}{\mathbf{GL}}

\newcommand{\ODD}{\mathbf{O}(D,D)}

\newcommand{\etaodd}{{\cJ}}

\newcommand{\Spint}{{\Spin(1,9)}}

\newcommand{\Ott}{\mathbf{O}(10,10)}

\newcommand{\SOt}{{\SO(1,9)}}
\newcommand{\oSOt}{{{\SO}(9,1)}}
\newcommand{\sot}{{\so(1,9)}}
\newcommand{\osot}{{{\so}(9,1)}}

\newcommand\SDFT{{\rm SDFT}}

\newcommand\cD{{\cal D}}
\newcommand\cE{{\cal E}}

\newcommand\cH{{\cal H}}

\newcommand\cJ{{\cal J}}

\newcommand\cL{{\cal L}}
\newcommand\cM{{\cal M}}
\newcommand\cN{{\cal N}}

\newcommand\cP{{\cal P}}

\newcommand\bcP{{\bar{\cP}}}

\newcommand\hcL{{\hat{\cal L}}}
\newcommand\hGamma{{\hat{\Gamma}}}

\newcommand\Phippp{{\Phi^{\prime\prime\prime}}{}}

\newcommand\cDs{\cD^{\star}}
\newcommand\cDh{\cD^{\sharp}}
\newcommand\cDhh{\cD^{\flat}}
\newcommand\Gammah{\Gamma^{\sharp}}
\newcommand\Gammahh{\Gamma^{\flat}}

\def\Gammas{\Gamma^{\star}}

\def\Gammao{\Gamma^{\scriptscriptstyle{0}}}
\def\Phio{\Phi^{\scriptscriptstyle{0}}}
\def\DOo{\DO^{\scriptscriptstyle{0}}}
\def\brPhio{\brPhi^{\scriptscriptstyle{0}}}

\def\cDo{\cD^{\scriptscriptstyle{0}}}
\def\hcD{\hat{\cD}}

\def\tx{\tilde{x}}

\def\bre{\bar{e}}
\def\brvare{\bar{\varepsilon}}
\def\breta{\bar{\eta}}

\def\brrho{\bar{\rho}}
\def\brpsi{\bar{\psi}}

\def\brp{{\bar{p}}}
\def\brq{{\bar{q}}}
\def\brr{{\bar{r}}}

\def\brPhi{{{\bar{\Phi}}}}
\def\brDelta{{{\bar{\Delta}}}}

\def\brF{\bar{F}}

\def\brV{{\bar{V}}}
\def\brP{{\bar{P}}}

\def\Tw{{T}}

\newcommand{\DO}{\mathbf{\nabla}}

\newcommand{\na}{{\nabla}}
\newcommand{\trd}{{\bigtriangledown}}

\renewcommand{\arraystretch}{1.25}       
\begin{document}
\begin{CJK}{UTF8}{mj}
\preprint{CERN-PH-TH/2011-278}
\title{Supersymmetric Double Field Theory: Stringy Reformulation of Supergravity}
\author{Imtak Jeon~(전임탁)${}^{\sharp\star}$,    ~Kanghoon Lee~(이강훈)${}^{\flat}$~ 
and~ Jeong-Hyuck Park~(박정혁)${}^{\star}$}

\affiliation{{~}\\
${}^{\sharp}$CERN, Theory Division CH-1211 Geneva 23, Switzerland\\
\mbox{${}^{\star}$Department of Physics,  Sogang University, Seoul 121-742, Korea}\\
\mbox{${}^{\flat}$Center for Quantum Spacetime,  Sogang University, Seoul 121-742, Korea}\\~\\
\textbf{\rm\small{imtak@sogang.ac.kr,~~kanghoon@sogang.ac.kr,~~park@sogang.ac.kr}}}


\begin{abstract}
We construct a supersymmetric extension of double field theory that realizes   the ten-dimensional Majorana-Weyl local supersymmetry.   In terms of  a stringy differential geometry we proposed earlier, our action  consists of five simple terms -- two bosonic  plus three fermionic  -- and manifests not only  diffeomorphism and  one-form gauge symmetry of $B$-field,   but also  $\Ott$ T-duality  as well as  a direct product of two local  Lorentz symmetries, ${\SOt\times\oSOt}$. A gauge fixing that  identifies  the double  local  Lorentz groups reduces our action  to the    minimal supergravity in ten dimensions.   
\end{abstract}
\maketitle
\end{CJK}

Without resorting to vector notation,   Maxwell's original equations consisted of twenty formulas. It was   the rotational  or  Lorentz symmetry that  reorganized  them into   four or two  compact   equations.   Recent developments in string theory indicate that  supergravity theories  -- at least those which  have stringy  origin  --  may undergo  a similar reformulation,  and be greatly simplified with  the renewed understanding  of their stringy   structure or T-duality.

T-duality is a genuine stringy effect such that  string theory effective actions or  ten-dimensional supergravities    should feature $\Ott$ structure~\cite{Buscher:1985kb,Buscher:1987sk,Buscher:1987qj,Giveon:1988tt}.  The $\Ott$ T-duality   can be manifestly  realized if  we  formally double  the spacetime dimension,   from  ten to twenty  with coordinates  $x^{\mu}\rightarrow y^{A}=(\tx_{\mu},x^{\nu})$~\cite{Tseytlin:1990nb,Tseytlin:1990va,Siegel:1993xq,Siegel:1993th},  and reformulate the ten-dimensional effective action in terms of twenty dimensional language \textit{i.e.}~tensors equipped with  $\Ott$ metric, 
\be
\etaodd_{AB}={{{{\left(\ba{cc}0&1\\1&0\ea\right)}}\,.}}
\label{ODDeta}
\ee
This kind of reformulation  was coined   Double Field Theory (DFT)~\cite{Hull:2009mi,Hull:2009zb,Hohm:2010jy,Hohm:2010pp}, and  has attracted  much attention in recent years~\cite{Kwak:2010ew,Jeon:2010rw,Hohm:2010xe,Jeon:2011kp,Hohm:2011ex,Berman:2011pe,Jeon:2011cn,Jeon:2011vx,Copland:2011yh,Thompson:2011uw,Hohm:2011zr,Hohm:2011dv,Albertsson:2011ux,Hohm:2011cp,Kan:2011vg,Aldazabal:2011nj,Geissbuhler:2011mx,Berman:2011kg,Copland:2011wx}.    In DFT,    as a  field theory counterpart to  the level matching condition of closed string theories,   
 the $\Ott$ d'Alembertian  operator must be trivial, acting on arbitrary fields as well as  their products,
\be
\ba{ll}
 \partial^{2}\Phi=\partial_{A}\partial^{A}\Phi\equiv 0\,,~~~&~~~
 \partial_{A}\Phi_{1}\partial^{A}\Phi_{2}\equiv0\,.
 \ea
 \label{constraint}
 \ee
 Hence locally, up to $\Ott$ rotation, all the  fields are independent of the dual coordinates, ${\frac{\partial~~}{\partial\tx_{\mu}}\equiv0}$, and   the theory is not truly doubled~\cite{Hohm:2010jy}. \\
\indent In a sense,   the $\Ott$ structure in DFT is  a ``meta-symmetry" rather than  a Noether symmetry,  since  only after  dimensional reductions   can it generate a Noether symmetry.  Another feature of DFT is that,  the  diffeomorphism and the one-form gauge symmetry of $B$-field  are 
 naturally unified  into what we may call ``double-gauge symmetry," as they are generated by the  generalized Lie derivative~\cite{Courant,Siegel:1993th,Gualtieri:2003dx,Grana:2008yw,Hohm:2010pp},
\be
\ba{l}
\hcL_{X}\Tw_{A_{1}\cdots A_{n}}:=X^{B}\partial_{B}\Tw_{A_{1}\cdots A_{n}}+\omega_{{\scriptscriptstyle{T\,}}}\partial_{B}X^{B}\Tw_{A_{1}\cdots A_{n}}\\
~~~+\sum_{i=1}^{n}(\partial_{A_{i}}X_{B}-\partial_{B}X_{A_{i}})\Tw_{A_{1}\cdots A_{i-1}}{}^{B}{}_{A_{i+1}\cdots  A_{n}}\,,
\ea
\label{tcL}
\ee
where   $\omega_{{\scriptscriptstyle{T\,}}}$ is the weight of $\Tw_{A_{1}\cdots A_{n}}$.  Since this differs from the ordinary Lie derivative, the underlying  differential geometry of DFT   is not Riemannian~\cite{Siegel:1993xq,Siegel:1993th,Courant,Gualtieri:2003dx,Grana:2008yw,Hitchin:2004ut,Hitchin:2010qz,Jeon:2010rw,Jeon:2011cn,Coimbra:2011nw,Jeon:2011vx,Hohm:2010xe} (see  \cite{Berman:2010is,Berman:2011cg} for  extensions  to $\cM$-theory). Namely, while doubling the spacetime dimension is sufficient to manifest the $\Ott$ structure,   the double-gauge symmetry (\ref{tcL}) calls for  novel mathematical treatment. \\
\indent In this paper, we construct a supersymmetric extension of double field theory   that manifests  simultaneously    $\Ott$ T-duality and  various gauge symmetries   listed  in Table \ref{TABsymmetry} including the double-gauge symmetry. In order to do so, we employ  the  stringy differential geometry  we developed earlier in Refs.\cite{Jeon:2010rw,Jeon:2011cn,Jeon:2011vx}.  Especially  we utilize  the   ``semi-covariant" derivatives proposed   therein. 
\begin{table}[H]
\begin{center}
{{\small{\begin{itemize}
\item $\Ott$ T-duality: \textit{Meta-symmetry}
\item Gauge symmetries 
\begin{enumerate}
\item Double-gauge symmetry
\begin{itemize}
\item Diffeomorphism
\item One-form gauge symmetry
\end{itemize}
\item Local    Lorentz  symmetries, ${\SOt\times\oSOt}$
\item Local  Majorana-Weyl supersymmetry
\end{enumerate}
\end{itemize}}}}
\caption{T-duality and gauge symmetries in super DFT.  }
\label{TABsymmetry}
\end{center}
\end{table}
The  supersymmetric DFT action  we construct  below, Eq.(\ref{SDFTL}),  reformulates  the ten-dimensional  minimal, \textit{i.e.~}${\cN=1}$  supergravity  into five simple terms, while doubling the local Lorentz symmetries.   For  a related  superspace analysis  we refer readers   to  an earlier work by Siegel~\cite{Siegel:1993th}.
\section*{Setup}
Our supersymmetric extension  is  minimal such that the field contents include  the DFT-dilaton, 
double-vielbeins, DFT-dilatino and gravitino, as well as  a local supersymmetry parameter,  
\be
\ba{llllll}
d\,,~~&~~
 V_{Ap}\,,~~&~~\brV_{B\brq}\,,~~&~~\rho^{\alpha}\,,~~&~~\psi_{\brp}^{\,\alpha}\,,~~&~~\varepsilon^{\alpha}\,.
\ea
\label{fieldcontent}
\ee
Their weights are trivial except the DFT-dilaton, $d$, as it is  related to the conventional  string dilaton, $\phi$, through $e^{-2d}=\sqrt{-g}e^{-2\phi}$~\cite{Hohm:2010jy}, such that  $e^{-2d}$ has  weight  unity and 
\be
\hcL_{X}d:=-\half e^{2d}\hcL_{X}\!\left(e^{-2d}\right)=X^{A}\partial_{A}d-\half\partial_{A}X^{A}\,.
\ee
\indent Every field in (\ref{fieldcontent}) is  covariant under all the bosonic symmetries in Table \ref{TABsymmetry}. The   indices of each field   denote    the  relevant symmetry  representations,  as summarized in Table \ref{TABindices}.
\begin{table}[H]
\begin{center}
\begin{tabular}{||c|c|c||}
\hline
Index~&~Representation~&~Metric\\
\hline
$A,B,\cdots$~&~$\left.\ba{c}\Ott\\\mbox{double-gauge}\!\!\ea\right\}$\,vector~&$\cJ_{AB}$ in Eq.(\ref{ODDeta})\\
$p,q,\cdots$~&~$\SOt$  vector~&$\eta_{pq}=\mbox{diag}(-++\cdots+)$ \\
$\brp,\brq,\cdots$~&~$\oSOt$  vector~&$\breta_{\brp\brq}=\mbox{diag}(+--\cdots-)$ \\
$\alpha,\beta,\cdots$~&~$\Spint$  spinor~&$C_{\alpha\beta}$  in Eq.(\ref{CcM})\\
\hline
\end{tabular}
\caption{Indices for each symmetry representation and the relevant    metrics that raise or lower the positions of them. }
\label{TABindices}
\end{center}
\end{table}
With the real $\SOt$ gamma matrices, $(\gamma^{p})^{\alpha}{}_{\beta}$,
the charge conjugation matrix, $C_{\alpha\beta}$, satisfies
\be
(C\gamma^{p_{1}p_{2}\cdots p_{n}})_{\alpha\beta}=-(-1)^{n(n+1)/2}(C\gamma^{p_{1}p_{2}\cdots p_{n}})_{\beta\alpha}\,,
\label{CcM}
\ee
and defines the conjugated  spinors, $\brpsi_{\brp\alpha}=\psi_{\brp}^{\,\beta}C_{\beta\alpha}$, 
$\brrho_{\alpha}=\rho^{\beta} C_{\beta\alpha}$, \textit{etc.}
All the spinors are  taken to be $\Ott$ singlet and  Majorana-Weyl,  
possessing     definite chiralities.  With $\gamma^{(10)}=\gamma^{012\cdots 9}$ they obey 
\be
\ba{lll}
\gamma^{(10)}\psi_{\brp}=+\psi_{\brp}\,,~&~~\gamma^{(10)}\rho=-\rho\,,~&~~
\gamma^{(10)}\varepsilon=+\varepsilon\,.
\ea
\ee
The   double-vielbein satisfies    the defining properties~\cite{Jeon:2011cn},
\be
\ba{ll}
V_{Ap}V^{A}{}_{q}=\eta_{pq}\,,~~&~~V_{Ap}\brV^{A}{}_{\brq}=0\,,\\
\brV_{A\brp}\brV^{A}{}_{\brq}=\breta_{\brp\brq}\,,~~&~~V_{Ap}V_{B}{}^{p}+\brV_{A\brp}\brV_{B}{}^{\brp}=\cJ_{AB}\,.
\ea
\label{DEFNG}
\ee
Hence it generates a pair of  rank-two  projections~\cite{Jeon:2010rw},
\be
\ba{ll}
P_{AB}:=V_{A}{}^{p}V_{Bp}\,,~~&~~\brP_{AB}:=\brV_{A}{}^{\brp}\brV_{B\brp}\,,
\ea
\ee
that are symmetric, orthogonal and complementary, as
\be
\ba{ll}
P_{AB}=P_{BA}\,,~~&~~\brP_{AB}=\brP_{BA}\,,\\
P_{A}{}^{B}P_{B}{}^{C}=P_{A}{}^{C}\,,~~&~~\brP_{A}{}^{B}\brP_{B}{}^{C}=\brP_{A}{}^{C}\,,\\
P_{A}{}^{B}\brP_{B}{}^{C}=0\,,~~&~~P_{A}{}^{B}+\brP_{A}{}^{B}=\delta_{A}{}^{B}\,.
\ea
\label{symP2}
\ee
Further they give a pair of  rank-six,  symmetric and traceless  projections~\cite{Jeon:2011cn},
\be
\ba{l}
\cP_{CAB}{}^{DEF}:=P_{C}{}^{D}P_{[A}{}^{[E}P_{B]}{}^{F]}+\textstyle{\frac{2}{9}}P_{C[A}P_{B]}{}^{[E}P^{F]D}\,,\\
\bcP_{CAB}{}^{DEF}:=\brP_{C}{}^{D}\brP_{[A}{}^{[E}\brP_{B]}{}^{F]}+\textstyle{\frac{2}{9}}\brP_{C[A}\brP_{B]}{}^{[E}\brP^{F]D}\,.
\ea
\label{P6}
\ee

\indent We are now ready to recall  the  three differential operators from \cite{Jeon:2010rw,Jeon:2011cn,Jeon:2011vx} and  generalize  them to include  `torsion',
\be 
\ba{l}
\na_{A}=\partial_{A}+\Gamma_{A}\,,\\
D_{A}=\partial_{A}+\Phi_{A}+\brPhi_{A}\,,\\
\cD_{A}=\partial_{A}+\Gamma_{A}+\Phi_{A}+\brPhi_{A}\,.
\ea
\label{cDA}
\ee
The first, $\na_{A}$, is the  semi-covariant derivative   for the double-gauge symmetry we developed in~\cite{Jeon:2010rw,Jeon:2011cn},
\be
\ba{ll}
\DO_{C}\Tw_{A_{1}A_{2}\cdots A_{n}}
\!:=\!&\!\partial_{C}\Tw_{A_{1}A_{2}\cdots A_{n}}-\omega_{{\scriptscriptstyle{T\,}}}\Gamma^{B}{}_{BC}\Tw_{A_{1}A_{2}\cdots A_{n}}\\
{}&\!\!\!+
\sum_{i=1}^{n}\,\Gamma_{CA_{i}}{}^{B}\Tw_{A_{1}\cdots A_{i-1}BA_{i+1}\cdots A_{n}}\,.
\ea
\label{semi-covD}
\ee
\indent The second, $D_{A}$, is a covariant derivative for the pair of  local Lorenz symmetries having  the connections,  $\Phi_{A}$ and  $\brPhi_{A}$  for $\SOt$ and $\oSOt$ respectively~\cite{Jeon:2011vx}.  \\
\indent The last,   $\cD_{A}$,  is the ``master" derivative  combining    $\na_{A}$ and $D_{A}$~\cite{Jeon:2011vx}. 
As for the unifying description of the  closed string massless bosonic sector, it annihilates  all the  bosonic fields in (\ref{fieldcontent}),
\be
\ba{l}
\cD_{A}V_{Bp}=\partial_{A}V_{Bp}+\Gamma_{AB}{}^{C}V_{Cp}+\Phi_{Ap}{}^{q}V_{Bq}=0\,,\\
\cD_{A}\brV_{B\brp}=\partial_{A}\brV_{B\brp}+\Gamma_{AB}{}^{C}\brV_{C\brp}+\brPhi_{A\brp}{}^{\brq}\brV_{B\brq}=0\,,\\
\cD_{A}d:=-\half e^{2d}\na_{A}\!\left(e^{-2d}\right)=\partial_{A}d+\half\Gamma^{B}{}_{BA}=0\,,
\ea
\label{ANN}
\ee
and also all the ``constants,"  $\cJ_{AB}$, $\eta_{pq}$, $\breta_{\brp\brq}$, $C_{\alpha\beta}$,  $(\gamma^{p})^{\alpha}{}_{\beta}$.\\
The connections are all skew-symmetric and  related to each other, from (\ref{DEFNG}), (\ref{ANN}),   through
\be
\ba{l}
\Phi_{Apq}=-\Phi_{Aqp}=V^{B}{}_{p}\na_{A}V_{Bq}\,,\\
\brPhi_{A\brp\brq}=-\brPhi_{A\brq\brp}=\brV^{B}{}_{\brp}\na_{A}\brV_{B\brq}\,,\\
\Gamma_{ABC}=-\Gamma_{ACB}=V_{B}{}^{p}D_{A}V_{Cp}+\brV_{B}{}^{\brp}D_{A}\brV_{C\brp}\,,
\ea
\label{c3}
\ee
such that they   assume  the following  most general forms,
\be
\ba{ll}
\Phi_{Apq}=\Phio_{Apq}+\Delta_{Apq}\,,~&~\brPhi_{A\brp\brq}=\brPhio_{A\brp\brq}+\brDelta_{A\brp\brq}\,,\\
\multicolumn{2}{c}{\Gamma_{CAB}=\Gammao_{CAB}+\Delta_{Cpq}V_{A}{}^{p}V_{B}{}^{q}+\brDelta_{C\brp\brq}\brV_{A}{}^{\brp}\brV_{B}{}^{\brq}\,.}
\ea
\label{PhibrPhi}
\ee
Here, from  \cite{Jeon:2011cn},
\be
\ba{l}
\Gammao_{CAB}=2\!\left(P\partial_{C}P\brP\right)_{[AB]}\\
~~~+2\left({{\brP}_{[A}{}^{D}{\brP}_{B]}{}^{E}}-{P_{[A}{}^{D}P_{B]}{}^{E}}\right)\partial_{D}P_{EC}\\
~~~-\textstyle{\frac{4}{9}}\left(\brP_{C[A}\brP_{B]}{}^{D}+P_{C[A}P_{B]}{}^{D}\right)\!\left(\partial_{D}d+(P\partial^{E}P\brP)_{[ED]}\right),
\ea
\label{Gammao}
\ee
and,  with  the corresponding derivative, $\DOo_{A}=\partial_{A}+\Gammao_{A}$,
\be
\ba{ll}
\Phio_{Apq}=V^{B}{}_{p}\DOo_{A} V_{Bq}\,,
~~&~~\brPhio_{A\brp\brq}=\brV^{B}{}_{\brp}\DOo_{A}\brV_{B\brq}\,.
\ea
\label{Phio}
\ee

As $\Gammao_{ABC}=\Gammao_{A[BC]}$  in (\ref{Gammao}) is  the unique  connection  that  further obeys~\cite{Jeon:2011cn,Jeon:2011vx},
\be
\ba{ll}
\Gammao_{[ABC]}=0\,,~&~
(\cP+\bcP)_{CAB}{}^{DEF}\Gammao_{DEF}=0\,,
\ea
\ee
$\Gammao_{A}$, $~\Phio_{B}$ ~and~  $\brPhio_{C}~$ correspond to  the ``minimal" or ``torsionless" connections. The extra  covariant  pieces, $\Delta_{Apq}=-\Delta_{Aqp}$ and $\brDelta_{A\brp\brq}=-\brDelta_{A\brq\brp}$,    then can be viewed as  torsion~\cite{Jeon:2011vx},  being subject to 
\be
\ba{ll}
\Delta_{Apq}V^{Ap}=0\,,~~&~~\brDelta_{A\brp\brq}\brV^{A\brp}=0\,,
\ea
\ee
which are necessary to maintain $\cD_{A}d=0$. As is the case in ordinary  supergravities,     the torsion can be constructed from the bi-spinorial  objects. Specifically,  in the present  work  we set
\be
\ba{ll}
\Gamma_{ABC}=&\Gammao_{ABC}+i\textstyle{\frac{1}{3}}\brrho\gamma_{ABC}\rho+i\textstyle{\frac{1}{3}}\brpsi^{\brp}\gamma_{ABC}\psi_{\brp}\\
{}&-2i\brrho\gamma_{BC}\psi_{A}-4i\brpsi_{B}\gamma_{A}\psi_{C}\,.
\ea
\label{Gamma}
\ee
Further, hereafter,  for simplicity we  put 
\be
\ba{ll}
\psi_{A}:=\brV_{A}{}^{\brp}\psi_{\brp}\,,~~~&~~~\gamma^{A}:=V^{A}{}_{p}\gamma^{p}\,,\\
\cD_{p}:=V^{A}{}_{p}\cD_{A}\,,~~~&~~~\cD_{\brp}:=\brV^{A}{}_{\brp}\cD_{A}\,,
\ea
\ee
such  that $\brV^{A}{}_{\brp}\psi_{A}=\psi_{\brp}\,$   and $\,\left\{\gamma^{A},\gamma^{B}\right\}=2P^{AB}$, \textit{etc}.\\
\indent From $[\cD_{A},\cD_{B}]V_{Cp}=0$,   $\,[\cD_{A},\cD_{B}]\brV_{C\brp}=0$,  
the usual  curvatures of the three connections, $\Gamma_{A}$, $\Phi_{A}$, $\brPhi_{A}$,
\be
\ba{l}
R_{CDAB}=\partial_{A}\Gamma_{BCD}+\Gamma_{AC}{}^{E}\Gamma_{BED}-({{A\,\leftrightarrow\, B}})\,,\\
F_{ABpq}=\partial_{A}\Phi_{Bpq}-\partial_{B}\Phi_{Apq}+\Phi_{Apr}\Phi_{B}{}^{r}{}_{q}-\Phi_{Bpr}\Phi_{A}{}^{r}{}_{q}\,,\\
\brF_{AB\brp\brq}=\partial_{A}\brPhi_{B\brp\brq}-\partial_{B}\brPhi_{A\brp\brq}+\brPhi_{A\brp\brr}\brPhi_{B}{}^{\brr}{}_{\brq}-\brPhi_{B\brp\brr}\brPhi_{A}{}^{\brr}{}_{\brq}\,,
\ea
\ee 
satisfy  the following relation, 
\be
R_{ABCD}=F_{CDpq}V_{A}{}^{p}V_{B}{}^{q}+\brF_{CD\brp\brq}\brV_{A}{}^{\brp}\brV_{B}{}^{\brq}\,.
\label{RFF}
\ee
However, they are not  double-gauge covariant~\cite{Jeon:2011cn}. Covariant quantities are achievable 
if we define~\cite{Jeon:2011cn}
\be
S_{ABCD}:=\half\left(R_{ABCD}+R_{CDAB}-\Gamma^{E}{}_{AB}\Gamma_{ECD}\right)\,,
\ee
satisfying, with  $\Gamma_{ABC}=\Gammao_{ABC}+\Lambda_{ABC}$ (\ref{PhibrPhi}),  (\ref{Gammao}),
\be
\ba{l}
S_{ABCD}=S^{\scriptscriptstyle{0}}_{ABCD}+\cDo_{[A}\Lambda_{B]CD}+\cDo_{[C}\Lambda_{D]AB}\\
~~~~~+\Lambda_{D[A}{}^{E}\Lambda_{|C|B]E}+\Lambda_{B[C}{}^{E}\Lambda_{|A|D]E}-\half\Lambda^{E}{}_{AB}\Lambda_{ECD}\,.
\ea
\ee 
Examples of the covariant quantities   include 
\be
\ba{c}
P^{AB}P^{CD}S_{ACBD}\,,~~~~~~~\brP^{AB}\brP^{CD}S_{ACBD}\,,\\
S_{p\brq}+2i\brpsi^{A}\gamma_{p}\cD_{A}\psi_{\brq}-2i\brpsi_{\brq}\cD_{p}\rho\,,\\
\gamma^{A}\cD_{A}\rho\,,~~~~~\gamma^{A}\cD_{A}\psi_{\brp}\,,~~~~~\cD_{\brp}\rho\,,~~~~~\cD_{A}\psi^{A}\,,\\
\brpsi^{A}\gamma_{p}(\cD_{A}\psi_{\brq}-\half\cD_{\brq}\psi_{A})\,,
 \ea
 \label{EXCOV}
 \ee
 where $S_{p\brq}=V^{A}{}_{p}\brV^{B}{}_{\brq}S_{AB}$ and  $S_{AB}=S_{ACB}{}^{C}$. This generalizes our earlier results~\cite{Jeon:2011cn,Jeon:2011vx} to the   torsionful connection~(\ref{Gamma}).
 
\section*{Supersymmetric DFT Lagrangian} 
The supersymmetric double field theory  Lagrangian we construct in this work consists of   five terms (\textit{cf.}~\cite{Siegel:1993th}):
\be
\ba{l}
\cL_{\SDFT}=e^{-2d}\Big[\textstyle{\frac{1}{8}}\left(P^{AB}P^{CD}-\brP^{AB}\brP^{CD}\right)S_{ACBD}\\
{}~~~~~~~~~~~+i\half\brrho \gamma^{A}\cDs_{A}\rho
+i\brpsi^{A}\cDs_{A}\rho+i\half\brpsi^{B}\gamma^{A}\cDs_{A}\psi_{B}\Big]\,,
\ea
\label{SDFTL}
\ee
where, with (\ref{Gamma}), $\cDs_{A}$ is defined  by its own connection,
\be
\ba{ll}
\Gammas_{ABC}=&\Gamma_{ABC}-i\textstyle{\frac{11}{96}}\brrho\gamma_{ABC}\rho-i\textstyle{\frac{5}{24}}\brpsi^{\brp}\gamma_{ABC}\psi_{\brp}\\
{}&+i\textstyle{\frac{5}{4}}\brrho\gamma_{BC}\psi_{A}+2i\brpsi_{B}\gamma_{A}\psi_{C}\,.
\label{Gammas}
\ea
\ee
From (\ref{EXCOV}) and \cite{Jeon:2011cn,Jeon:2011vx}, each term in the Lagrangian   is   invariant under  all the bosonic symmetries listed in Table~\ref{TABsymmetry},  while the whole  Lagrangian  is supersymmetric,  up to the strong  level matching constraint (\ref{constraint}),  under
\be
\ba{cl}
\delta_{\varepsilon} d&=\,i\half\brvare\rho\,,\\
\delta_{\varepsilon} V_{Ap}&=\,i\brvare\gamma_{p}\psi_{A}\,,\\
\delta_{\varepsilon} \brV_{A\brp}&=\,-i\brvare\gamma_{A}\psi_{\brp}\,,\\
\delta_{\varepsilon}\rho&=\,-\gamma^{A}\hcD_{A}\varepsilon\,,\\
\delta_{\varepsilon}\psi_{\brp}&=\,\brV^{A}{}_{\brp}\hcD_{A}\varepsilon-i\quarter(\brrho\psi_{\brp})\varepsilon-i\half(\brvare\rho)\psi_{\brp}\,,
\ea
\label{SUSY}
\ee
where, again with (\ref{Gamma}), $\hcD_{A}$ is set by another  connection,
\be
\hGamma_{ABC}=\Gamma_{ABC}-i\textstyle{\frac{17}{48}}\brrho\gamma_{ABC}\rho-i\textstyle{\frac{1}{4}}\brpsi^{\brp}\gamma_{ABC}\psi_{\brp}+i\textstyle{\frac{5}{2}}\brrho\gamma_{BC}\psi_{A}\,.
\label{hGamma}
\ee

Under arbitrary variations of all the fields, we get
\be
\ba{l}
\delta P_{AB}=-\delta\brP_{AB}=2\delta V_{(A}{}^{p}V_{B)p}\,,\\
\delta V_{Ap}=\delta V_{Bp}\brP^{B}{}_{A}+\delta V_{B[p}V^{B}{}_{q]}V_{A}{}^{q}\,,\\
\delta\psi_{A}=\left(\delta\psi_{\brp}+\psi_{\brq}\delta\brV_{B}{}^{\brq}\brV^{B}{}_{\brp}\right)\brV_{A}{}^{\brp}
-\psi_{B}\delta V^{B}{}_{p}V_{A}{}^{p}\,,\\
\delta\Phi_{Apq}=\cD_{A}(V^{B}{}_{p}\delta V_{Bq})+V^{B}{}_{p}V^{C}{}_{q}\delta\Gamma_{ABC}\,,\\
\delta\brPhi_{A\brp\brq}=\cD_{A}(\brV^{B}{}_{\brp}\delta \brV_{B\brq})+\brV^{B}{}_{\brp}\brV^{C}{}_{\brq}\delta\Gamma_{ABC}\,,\\
\delta S_{ABCD}\!=\!\cD_{[A}\delta\Gamma_{B]CD}-\!\textstyle{\frac{3}{2}}
\Gamma_{[EAB]}\delta\Gamma^{E}{}_{CD}+\![{\scriptstyle(A,B)\,\leftrightarrow\, (C,D)}],
\ea
\label{useful}
\ee
and,  the Lagrangian transforms   up to total derivatives (\,$\cong$\,)  as
\be
\ba{l}
\delta\cL_{\SDFT}\cong-2\delta d\times\cL_{\SDFT}\,+\,\delta\Gamma_{ABC}\times 0\\
~~~~~~~+\half e^{-2d} \delta V^{Bp}\brV_{B}{}^{\brq}\\
~~~~~~~~~~~\times\Big(S_{p\brq}-2i\brpsi_{\brq}\hcD_{p}\rho
+i\brpsi^{A}\gamma_{p}\cDhh_{\brq}\psi_{A}+i\brrho\gamma_{p}\cDh_{\brq}\rho\Big)\\
~~~~~~~+ie^{-2d}\left(\delta\brrho-\textstyle{\frac{1}{4}}\delta V_{Bq}\brrho\gamma^{Bq}\right)\times
\Big(\gamma^{A}\cDh_{A}\rho-\hcD_{A}\psi^{A}\Big)\\
~~~~~~~+ie^{-2d}\Big(\delta\brpsi^{\brp}+\brpsi^{\brq}\delta\brV_{B\brq}\brV^{B\brp}
-\textstyle{\frac{1}{4}}\delta V_{Bq}\brpsi^{\brp}\gamma^{Bq}\Big)\\
~~~~~~~~~~~~~~~~~~~~~~\times\Big(\hcD_{\brp}\rho+\gamma^{A}\cDhh_{A}\psi_{\brp}\Big)\,.
\ea
\label{opf}
\ee
From this, covariant,   four sorts  of equations of motion  (two bosonic and two fermionic) can be readily read off.  Here we let   for  $\cDh_{A}$,  $\cDhh_{A}$ in the fermionic equations of motion,
\be
\ba{l}
\Gammah_{ABC}=\Gamma_{ABC}-i\textstyle{\frac{31}{96}}\brpsi^{\brr}\gamma_{ABC}\psi_{\brr}
+i\textstyle{\frac{17}{24}}\brrho\gamma_{BC}\psi_{A}\,,\\
\Gammahh_{ABC}=\Gamma_{ABC}-i\textstyle{\frac{31}{96}}
\brrho\gamma_{ABC}\rho-i\textstyle{\frac{5}{12}}\brpsi^{\brr}\gamma_{ABC}\psi_{\brr}\\
~~~~~~~~~~~~+i\textstyle{\frac{1}{2}}\brrho\gamma_{BC}\psi_{A}+4i\brpsi_{B}\gamma_{A}\psi_{C}\,.
\ea
\ee
\indent  The 1.5 formalism that is familiar  in ordinary supergravities   holds in (\ref{opf}): the variation of the Lagrangian by $\delta\Gamma_{ABC}$, with (\ref{useful}),   identically vanishes for the solution (\ref{Gamma}), while  the equations of motion for the fermions can be obtained from  the fermionic sector only \textit{i.e.~}the last three terms in the Lagrangian~(\ref{SDFTL}), with (\ref{Gammas}).

\section*{DFT Supersymmetry algebra}
Starting from the supersymmetry transformation rule of  each DFT field  in Eq.(30), through straightforward yet somewhat lengthy computations, we can obtain  the following supersymmetry commutator relations,  up to the strong   level matching constraint (\ref{constraint}),  
\be
\ba{l}
\left[\delta_{\varepsilon_{1}},\delta_{\varepsilon_{2}}\right]d\equiv
\hcL_{X_{3}}d\,,
\\
\left[\delta_{\varepsilon_{1}},\delta_{\varepsilon_{2}}\right]V_{Ap}\equiv
\hcL_{X_{3}}V_{Ap}+\delta_{\varepsilon_{3}}V_{Ap}+\Lambda_{pq}V_{A}{}^{q}\,,
\\
\left[\delta_{\varepsilon_{1}},\delta_{\varepsilon_{2}}\right]\brV_{A\brp}\equiv
\hcL_{X_{3}}\brV_{A\brp}+\delta_{\varepsilon_{3}}\brV_{A\brp}+\bar{\Lambda}_{\brp\brq}\brV_{A}{}^{\brq}\,,
\\
\left[\delta_{\varepsilon_{1}},\delta_{\varepsilon_{2}}\right]\rho\equiv\hcL_{X_{3}}\rho
+\delta_{\varepsilon_{3}}\rho+\quarter\Lambda_{pq}\gamma^{pq}\rho\\
~~~~~~~~~~~~~~~~-\half X_{3}^{p}\gamma_{p}\left(\gamma^{A}\cDh_{A}\rho-\hcD_{A}\psi^{A}\right),\\
\left[\delta_{\varepsilon_{1}},\delta_{\varepsilon_{2}}\right]\psi_{\brp}\equiv
\hcL_{X_{3}}\psi_{\brp}+\delta_{\varepsilon_{3}}\psi_{\brp}
+\quarter\Lambda_{pq}\gamma^{pq}\psi_{\brp}+\bar{\Lambda}_{\brp\brq}\psi^{\brq}\\
~~+\textstyle{\frac{1}{32}}\left(\textstyle{\frac{1}{5!}}Y^{mnpqr}\gamma_{mnpqr}
 -14X_{3}^{p}\gamma_{p}\right)\left(\hcD_{\brp}\rho+\gamma^{A}\cDhh_{A}\psi_{\brp}\right),
\ea
\label{SUSYCOM}
\ee
where, with 
\be
\ba{l}
Y^{mnpqr}=i\bar{\varepsilon}_{2}\gamma^{mnpqr}\varepsilon_{1}\,,\\
\Phi^{\prime}_{Apq}=\Phio_{Apq}+i\textstyle{\frac{1}{24}}\brrho\gamma_{Apq}\rho-i\textstyle{\frac{7}{24}}\brpsi^{\brr}\gamma_{Apq}\psi_{\brr}\,,
\ea
\ee
the parameters  are given by
\be
\ba{l}
X_{3}^{A}=i\bar{\varepsilon}_{2}\gamma^{A}\varepsilon_{1}\,,\\
\varepsilon_{3}=i\half\left[(\bar{\varepsilon}_{2}\gamma^{p}\varepsilon_{1})\gamma_{p}\rho
+(\bar{\rho}\varepsilon_{2})\varepsilon_{1}-(\bar{\rho}\varepsilon_{1})\varepsilon_{2}\right]\,,\\
\Lambda_{pq}=2\hcD_{[p}X_{3}{}_{q]}-i\textstyle{\frac{1}{48}}
\brpsi^{\brp}\gamma^{lmn}\psi_{\brp}Y_{lmnpq}+\Phi^{\prime}_{Apq}X_{3}^{A}\,,\\
\bar{\Lambda}_{\brp\brq}=\left(\brPhio_{A\brp\brq}-i\brpsi_{\brp}\gamma_{A}\psi_{\brq}\right)X_{3}^{A}\,.
\ea
\label{PARA}
\ee
In particular, $\Lambda_{pq}{=-\Lambda_{qp}}$ and $\bar{\Lambda}_{\brp\brq}{=-\bar{\Lambda}_{\brq\brp}}$ correspond to the  $\sot$ and $\osot$  local Lorentz symmetry parameters respectively. Further, since $X_{3}^{p}C\gamma_{p}$ and $C(\textstyle{\frac{1}{5!}}Y^{mnpqr}\gamma_{mnpqr}
 -14Y^{p}\gamma_{p})$ are symmetric,  the terms that are proportional to the fermionic equations of motion in (\ref{SUSYCOM}) correspond to the fermionic ``trivial" gauge symmetry~\cite{Henneaux:1992ig}.

 Identifying  the   common  symmetry parameters  in (\ref{PARA})  on the right hand side of each line in (\ref{SUSYCOM}) provides  a nontrivial consistency check.  For this,  note  also   identically,
\be
\delta_{\varepsilon_{3}}d=i\half\bar{\varepsilon}_{3}\rho=0\,.
\ee
Therefore,  the commutator of DFT supersymmetry transformations (\ref{SUSY}) closes up to the  strong  level matching constraint  and every   gauge symmetry   listed  in Table~\ref{TABsymmetry},  as well as the  fermionic equations of motion,
\be
[\delta_{\varepsilon_{1}},\delta_{\varepsilon_{2}}]\equiv\hcL_{X_{3}}+\delta_{\varepsilon_{3}}+\delta_{\sot}+\delta_{\osot}+\delta_{\rm{{trivial}}}\,.
\label{parameters}
\ee
\newpage

\section*{Comments}   
\indent From (\ref{Gammao}), (\ref{Gamma}), (\ref{RFF}),  up to  the strong level matching constraint (\ref{constraint}),   we obtain
\[
\ba{c}
{P^{AB}\brP^{CD}S_{ACBD}=-\half P^{AB}\brP^{CD}\Gamma^{E}{}_{AB}\Gamma_{ECD}\equiv 0\,,}\\
{P^{AB}P^{CD}S_{ACBD}\equiv P^{AB}S_{AB}}\,,\\
{\brP^{AB}\brP^{CD}S_{ACBD}\equiv\brP^{AB}S_{AB}}\,.
\ea
\]
Hence,  the bosonic part of  the Lagrangian~(\ref{SDFTL}) may reduce to the single term,  $\textstyle{\frac{1}{8}}\cH^{AB}S_{AB}$ that was previously suggested  in \cite{Jeon:2011cn} with the so-called  generalized metric, $\cH_{AB}=P_{AB}-\brP_{AB}$.  However, the expression  in (\ref{SDFTL}) appears  more directly relevant to the $1.5$ formalism (\ref{opf}).\\

\indent The double-vielbeins, $V_{Ap}$, $\brV_{A\brp}$ (\ref{DEFNG}),  admit  explicit parametrization in terms of the  Kalb-Ramond  $B$-field and a pair of zehnbeins  corresponding  to  the common   spacetime metric, $e_{\mu}{}^{p}e_{\nu}{}^{q}\eta_{pq}=-\bre_{\mu}{}^{{\brp}}\bre_{\nu}{}^{\brq}\breta_{\brp\brq}=g_{\mu\nu}$,   in an   $\Ott$ covariant   manner~\cite{Jeon:2011cn,Jeon:2011vx}. 
Gauge fixing  the two zehnbeins    equal to each other breaks  $\Ott$ to ${\mathbf{O}(10)\rtimes\GL(10)}$~\cite{Jeon:2011cn} and the pair of local Lorentz symmetries to a single one. Further, as shown  in detail  in the Appendix,  it  reduces our supersymmetric DFT  to  the  ten-dimensional ${\cN=1}$  supergravity of   eleven-dimensional origin~\cite{LPTENS-78-10}~(\textit{c.f.~}\cite{Chamseddine:1980cp,Bergshoeff:1981um}).   This result seems to   suggest  that  a generic  supergravity theory is an $\ODD$ and hence double local Lorentz   broken double field theory.  \\

\indent  The supersymmetric completion   of Refs.\cite{Hohm:2011zr,Hohm:2011dv} for  type IIA/IIB supergravity  remains  as a  future work.\\
{}~\\
\indent\textit{Acknowledgements}:  We wish to thank  Neil Copland and  Dimitris Tsimpis    for  useful  comments.  The work was supported by the National Research Foundation of Korea\,(NRF) grants  funded by the Korea government\,(MEST) with the Grant No.  2005-0049409 (CQUeST)  and Grant No.  2010-0002980.  The work by IJ is partially supported by NRF though the Korea-CERN theory collaboration.\\
~\\
\indent \textit{Note added}: After our submission, Ref.\cite{Hohm:2011nu} appeared in arXiv which also addresses   the supersymmetrization of DFT, yet  up to the quadratic order in fermions. It differs  in detail  from our full order analysis.\\
~\\

 ~\\
 ~\\ 
  
\newpage

\appendix
\setlength{\jot}{9pt}                 
\renewcommand{\arraystretch}{2} 

\begin{widetext}
\section{Reduction from $11D$ to $10D$ supergravity\label{11D10D}}
Here after spelling out   the   $11D$ supergravity  by Cremmer, Julia and Scherk~\cite{LPTENS-78-10}, we  set up our ansatz  of  its  dimensional reduction to the $10D$ minimal supergravity. {Our ansatz  differs  in detail  from those in \cite{Chamseddine:1980cp,Bergshoeff:1981um}, and   is designed to produce the precise  $10D$  $\cN=1$ supergravity in string frame  with  which   our supersymmetric DFT  matches.}\\~\\
\subsection{$11D$ supergravity from  Ref.\cite{LPTENS-78-10}}
With the  eleven-dimensional  curved and flat vector indices,  $M,N,P,\cdots$ and $A,B,C,\cdots$ respectively,  the $11D$ supergravity action is
\be
\ba{ll}
\cL_{11D}=&\frac{E}{4\kappa^{2}}R(E,\omega)-i\frac{E}{2}\bar{\Psi}_{M}\Gamma^{MNP}D_{N}(\frac{\omega+\hat{\omega}}{2})\Psi_{P}-\frac{E}{48}F_{MNPQ}F^{MNPQ}\\
{}&+i\frac{E\kappa}{192}(\bar{\Psi}_{M}\Gamma^{MNPQRS}\Psi_{N}+12\bar{\Psi}^{P}\Gamma^{QR}\Psi^{S})(F_{PQRS}+\hat{F}_{PQRS})\\
{}&-\frac{2\kappa}{(144)^2}\epsilon^{M_{1}M_{2}M_{3}M_{4}N_{1}N_{2}N_{3}N_{4}P_{1}P_{2}P_{3}}F_{M_{1}M_{2}M_{3}M_{4}}F_{N_{1}N_{2}N_{3}N_{4}}A_{P_{1}P_{2}P_{3}}\,,
\ea\ee
where,   $D_{M}$ is a covariant derivative with respect to the $11D$ local Lorentz transformation only, such that  with the standard  Christoffel symbol, it satisfies
\be
\ba{ll}
D_{M}(\omega)E_{N}{}^{A}=\Gamma_{M}{}^{P}{}_{N}E_{P}{}^{A}\,,~~~&~~~
\Gamma_{M}{}^{P}{}_{N}=\left\{{}_{M}{}^{P}{}_{N}\right\}+h_{M}{}^{P}{}_{N}\,.
\ea
\ee
Further  we have
\be
\ba{l}
\omega_{MAB}=-\omega_{MBA}=\half\left[ E_{A}{}^{N}(\partial_{M}E_{NB}-\partial_{N}E_{MB}+E_{M}{}^{C}\partial_{P}E_{NC}E^{P}{}_{B})-(A\leftrightarrow B)\right]+h_{MAB}\,,\\
h_{MAB}=-h_{MBA}=i\kappa^2 \left(\quarter\Psi_{P}\Gamma_{MAB}{}^{PQ}\Psi_{Q}+\bar{\Psi}_{M}\Gamma_{[A}\Psi_{B]}+
\half\bar{\Psi}_{A}\Gamma_{M}\Psi_{B}  \right)\,,
\ea\ee
and
\be
\ba{l}
\hat{\omega}_{MAB}=\omega_{MAB}-i\frac{\kappa^2}{4}\bar{\Psi}_{P}\Gamma^{PQ}{}_{MAB}\Psi_{Q}\,,\\
\hat{F}_{MNPQ}=F_{MNPQ}-3i\kappa\bar{\Psi}_{[M}\Gamma_{NP}\Psi_{Q]}\,.
\ea
\ee
The supersymmetry transformations  are
\be
\ba{l}
\delta_{\cE} E^{A}{}_{M}=i\kappa\bar{\cE}\,\Gamma^{A}\Psi_{M}\,,\\
\delta_{\cE}\Psi_{M}=\frac{1}{\kappa}\hat{D}_{M}(\hat{\omega},\hat{F})\cE\,,\\
\delta_{\cE} A_{MNP}=i\frac{3}{2}\bar{\cE}\,\Gamma_{[MN}\Psi_{P]}\,,
\ea
\label{11DSUSY}
\ee
where 
\be
\hat{D}_{M}(\hat{\omega},\hat{F})=D_{M}(\hat{\omega})-\textstyle{\frac{\kappa}{144}}(\Gamma_{M}{}^{PQRS}-8\delta^{P}_{M}\Gamma^{QRS})\hat{F}_{PQRS}\,.
\ee
~\\~\\

\subsection{Ansatz of the reduction}
With the decomposition of the curved and  the flat $11D$ vector indices, 
\be
\ba{ll}
M=(\mu, 11)\,,~~~~&~~~~A=(a, z)\,,
\ea
\ee
after putting $\kappa=1$, our ansatz  of the reduction is as follows. For  the {elfbein} we set 
\be
E_{M}{}^{A}=\left(\ba{cc}e^{-\frac{1}{3}\phi}e_{\mu}{}^{a}\,&0\\0&\,e^{\frac{2}{3}\phi}\ea\right)\,,
\ee
and for the three-form gauge field, we put
\be
\ba{ll}
A_{\mu\nu\lambda}=0\,,~~~~&~~~~
A_{\mu\nu 11}=\frac{1}{2}B_{\mu\nu}\,.
\ea
\ee
Further for the fermions, we write
{\be
\ba{l}
\Psi_{a}=\frac{1}{6}2^{\frac{1}{4}}e^{\frac{1}{6}\phi}\left(5\psi_{a}-\gamma_{ab}\psi^{b}-\gamma_{a}\rho\right)\,,\\
\Psi_{z}=-\frac{1}{3}2^{\frac{1}{4}}e^{\frac{1}{6}\phi}\left(\rho+\gamma^{a}\psi_{a} \right)\,,\\
\cE=2^{-\frac{1}{4}}e^{-\frac{1}{6}\phi}\varepsilon\,,
\label{fermi}
\ea
\ee}
and impose the  chirality conditions,
\be
\ba{lll}
\gamma^{(10)}\psi_{a}=\psi_{a}\,,~~~&~~~~
\gamma^{(10)}\rho=-\rho\,.~~~&~~~~
\gamma^{(10)}\varepsilon=\varepsilon\,.
\ea
\ee
Finally, we  reduce the $11D$ supersymmetry~(\ref{11DSUSY}) to $10D$ $\cN=1$ supersymmetry,  by adding  an $\sot$ local Lorentz transformation,  parametrized by $\Lambda^{\prime}_{ab}=i\textstyle{\frac{1}{6}}\bar{\varepsilon}\gamma_{ab}(\rho+\gamma^{c}\psi_{c})$,
\be
\delta^{11D}_{\cE}+\delta_{\Lambda^{\prime}}\Rightarrow\delta^{10D}_{\varepsilon}\,.
\label{10DSUSYpre}
\ee
~\\

\subsection{$10D$ ${\cN=1}$ supergravity}
The resulting $10D$ ${\cN=1}$ supergravity action is, after heavy usage of the Fierz identities (\ref{r4}), (\ref{rrpp}), (\ref{p2}), (\ref{p3}), (\ref{p4}),  
\be
\ba{ll}
\cL_{10D}= e\times e^{-2\phi}\Big[&\!\!R+4\partial_{\mu}\phi\partial^{\mu}\phi-\textstyle{\frac{1}{12}}H_{\lambda\mu\nu}H^{\lambda\mu\nu}\\
{}&~+i{2\sqrt{2}}\bar{\rho}\gamma^{m}( \partial_{m} \rho + \frac{1}{4} \omega_{m n p} \gamma^{n p} \rho + \frac{1}{24} H_{m n p} \gamma^{n p} \rho)\\
{}&~-i{4\sqrt{2}}\bar{\psi}^{p}(\partial_{p} \rho + \frac{1}{4} \omega_{p q r} \gamma^{q r} \rho + \frac{1}{8} H_{p q r} \gamma^{q r} \rho)\\
{}&~-i{2\sqrt{2}}\bar{\psi}^{p} \gamma^{m}( \partial_{m} \psi_{p} + \frac{1}{4} \omega_{m n p} \gamma^{n p} \psi_{p} +\omega_{m pq} \psi^{q} + \frac{1}{24} H_{m n p} \gamma^{n p} \psi_{p} - \half H_{m p q} \psi^{q}) \\
{}&~+\frac{1}{24}(\bar{\psi}^{q}\gamma_{mnp}\psi_{q})(\bar{\psi}^{r}\gamma^{mnp}\psi_{r})-\frac{1}{48}(\bar{\psi}^{q}\gamma_{mnp}\psi_{q})(\bar{\rho}\gamma^{mnp}\rho)~\Big]\,,
\ea
\label{10DSUGRA}
\ee
of which  the   ${\cN=1}$ supersymmetry  is, from (\ref{10DSUSYpre}),  given by
\be
\ba{l}
\delta_{\varepsilon} \phi = i \half \bar{\varepsilon} (\rho+\gamma^{a}\psi_{a}) \,, \\
\delta_{\varepsilon} e_{\mu}^{a} = i \bar{\varepsilon} \gamma^{a} \psi_{\mu}\,,\\
\delta_{\varepsilon} B_{\mu \nu} = -2i \bar{\varepsilon} \gamma_{[\mu} \psi_{\nu]}\,, \\
\delta_{\varepsilon} \rho = -\frac{1}{\sqrt{2}} \gamma^{a} (\partial_{a}\varepsilon + \frac{1}{4} \omega_{a b c} \gamma^{bc} \varepsilon + \frac{1}{24} H_{a b c} \gamma^{b c} \varepsilon - \partial_{a} \phi \varepsilon)\\
~~~~~~~~~+i \frac{1}{48} (\bar{\psi^{d}} \gamma_{a b c} \psi_{d}) \gamma^{a b c} \varepsilon +i \frac{1}{192} (\bar{\rho} \gamma_{a b c} \rho) \gamma^{a b c}\varepsilon\\
~~~~~~~~~+i\frac{1}{2}(\bar{\varepsilon}\gamma_{[a}\psi_{b]})\gamma^{ab}\rho\,,\\
\delta_{\varepsilon} \psi_{a} = \frac{1}{\sqrt{2}} (\partial_{a} \varepsilon + \frac{1}{4} \omega_{a b c} \gamma^{bc} \varepsilon + \frac{1}{8} H_{a b c} \gamma^{b c}\varepsilon) \\
~~~~~~~~~- i\frac{1}{2} (\bar{\rho}\varepsilon) \psi_{a} - i\frac{1}{4} (\bar{\rho} \psi_{a}) \varepsilon + i \frac{1}{8} (\bar{\rho} \gamma_{b c} \psi_{a}) \gamma^{b c} \varepsilon\\
~~~~~~~~~+i\frac{1}{2}(\bar{\varepsilon}\gamma_{[b}\psi_{c]})\gamma^{bc}\psi_{a} \,.
\ea
\label{10DSUSY}
\ee
Here  we set  $\partial_{p}=(e^{-1})_{p}{}^{\mu}\partial_{\mu}$ and  assume  the standard  spin connection,
$\omega_{\mu pq}=(e^{-1})_{p}{}^{\nu}\trd_{\mu}e_{\nu q}$,  with the  diffeomorphsim covariant derivative, $\trd_{\mu}$,  given by the torsionless, Christoffel symbol.\\~\\


\subsection{Matching the supersymmetric DFT and the $10D$ supergravity}
The   double-vielbein satisfies  the defining properties~(8),
\be
\ba{ll}
V_{Ap}V^{A}{}_{q}=\eta_{pq}\,,~~~&~~~V_{Ap}\brV^{A}{}_{\brq}=0\,,\\
\brV_{A\brp}\brV^{A}{}_{\brq}=\breta_{\brp\brq}\,,~~~&~~~V_{Ap}V_{B}{}^{p}+\brV_{A\brp}\brV_{B}{}^{\brp}=\cJ_{AB}\,,
\ea
\label{defV}
\ee
which are manifestly $\Ott$ covariant. Assuming that the upper half blocks of $V_{AP}$ and $\brV_{A\brp}$ are non-degenerate,  the  double-vielbein takes the following most general form~\cite{Jeon:2011cn,Jeon:2011vx}
\be
\ba{ll}
V_{Ap}=\textstyle{\frac{1}{\sqrt{2}}}{{\left(\ba{c} (e^{-1})_{p}{}^{\mu}\\(B+e)_{\nu p}\ea\right)}}\,,~
&~\brV_{A{\brp}}=\textstyle{\frac{1}{\sqrt{2}}}\left(\ba{c} (\bre^{-1})_{\brp}{}^{\mu}\\(B+\bre)_{\nu{\brp}}\ea\right)\,.
\ea
\label{Vform}
\ee
Here  $e_{\mu}{}^{p}$ and  $\bre_{\nu}{}^{{\brp}}$ are two copies of  the zehnbeins   corresponding  to  the same  spacetime metric,   
\be
e_{\mu}{}^{p}e_{\nu}{}^{q}\eta_{pq}=-\bre_{\mu}{}^{{\brp}}\bre_{\nu}{}^{\brq}\breta_{\brp\brq}=g_{\mu\nu}\,,
\label{sameg}
\ee
and $B_{\mu\nu}=-B_{\nu\mu}$ can be identified as  the Kalb-Ramond two-form gauge field.  We also set  in (\ref{Vform}),
\be
\ba{ll}
B_{\mu p}=B_{\mu\nu}(e^{-1})_{p}{}^{\nu}\,,~~~~&~~~~B_{\mu\brp}=B_{\mu\nu}(\bre^{-1})_{{\brp}}{}^{\nu}\,.
\ea
\ee
In particular, $(\bre^{-1}e)_{\brp}{}^{p}$ and $(e^{-1}\bre)_{p}{}^{\brp}$ are local Lorentz transformations, satisfying 
\be
\ba{l}
(\bre^{-1}e)_{\brp}{}^{p}(\bre^{-1}e)_{\brq}{}^{q}\eta_{pq}=-\breta_{\brp\brq}\,,\\
(e^{-1}\bre)_{p}{}^{\brp}(e^{-1}\bre)_{q}{}^{\brq}\breta_{\brp\brq}=-\eta_{pq}\,.
\ea
\label{ebreeta}
\ee

Since  (\ref{defV}) is manifestly $\Ott$ covariant and  the parametrization (\ref{Vform}) is quite   generic, the constraint (\ref{sameg}) is compatible with the  $\Ott$ structure.    For its explicit verification, we refer   (41), (42), (43) in \cite{Jeon:2011cn} or section 3.2 of \cite{Jeon:2011vx}.\\

Now, with  the explicit  parametrization (\ref{Vform}), from (31) in \cite{Jeon:2011cn} and (4.51) in \cite{Jeon:2011vx},  upon the strong  level matching constraint, $\tilde{\partial}\equiv 0$, our supersymmetric DFT Lagrangian (28) decomposes into three parts: genuine bosonic terms,   quadratic fermion terms,  and quartic fermion terms,
\be
\cL_{{\rm{SDFT}}}=\cL_{0}+\cL_{2}+\cL_{4}\,,
\ee
each of which reads explicitly,
\be
\ba{ll}
e^{2d}\cL_{0}&=\frac{1}{8}\left(P^{AB}P^{CD}-\bar{P}^{AB}\bar{P}^{CD}\right)S^{0}_{ABCD}\\
{}&\equiv\frac{1}{8}\left(R+4\Box\phi- 4\partial_{\mu}\partial^{\mu}\phi-\frac{1}{12}H_{\lambda\mu\nu}H^{\lambda\mu\nu}\right)\,,\\
e^{2d}\cL_{2}&=i\frac{1}{2}\bar{\rho}\gamma^{A}\cDo_{A}\rho+i\bar{\psi}^{A}\cDo_{A}\rho+i\frac{1}{2}\bar{\psi}^{B}\gamma^{A}\cDo_{A}\psi_{B}\\
{}&\equiv i\frac{1}{2\sqrt{2}}\bar{\rho}\gamma^{m}\left( \partial_{m} \rho + \frac{1}{4} \omega_{m n p} \gamma^{n p} \rho + \frac{1}{24} H_{m n p} \gamma^{n p} \rho  \right)+i\frac{1}{\sqrt{2}}\bar{\psi}^{\brp}\left(\partial_{\brp} \rho + \frac{1}{4} \omega_{\brp q r} \gamma^{q r} \rho + \frac{1}{8} H_{\brp q r} \gamma^{q r} \rho \right)\\
{}&~~~~+i\frac{1}{2\sqrt{2}}\bar{\psi}^{\brp} \gamma^{m} \left( \partial_{m} \psi_{\brp} + \frac{1}{4} \omega_{m n p} \gamma^{n p} \psi_{\brp} + \bar{\omega}_{m \brp\brq} \psi^{\bar{q}} + \frac{1}{24} H_{m n p} \gamma^{n p} \psi_{\brp} + \half H_{m \brp \brq} \psi^{\brq} 
\right)\,, \\
e^{2d}\cL_{4}&=\frac{1}{8}\left[\frac{1}{24}(\bar{\psi}^{D}\gamma_{ABC}\psi_{D})(\bar{\psi}^{E}\gamma^{ABC}\psi_{E})+\frac{1}{48}(\bar{\psi}^{D}\gamma_{ABC}\psi_{D})(\bar{\rho}\gamma^{ABC}\rho)\right]\\
{}&= \frac{1}{8}\left[\frac{1}{24}(\bar{\psi}^{\brp}\gamma_{mnp}\psi_{\brp})(\bar{\psi}^{\brq}\gamma^{mnp}\psi_{\brq})+\frac{1}{48}(\bar{\psi}^{\brp}\gamma_{mnp}\psi_{\brp})(\bar{\rho}\gamma^{mnp}\rho)
\right]
\,,
\ea
\label{L024}
\ee
where  $\partial_{p}=(e^{-1})_{p}{}^{\mu}\partial_{\mu}$,  $\,\partial_{\brp}=(\bre^{-1})_{\brp}{}^{\mu}\partial_{\mu}$, 
$\,\omega_{\mu pq}=(e^{-1})_{p}{}^{\nu}\trd_{\mu}e_{\nu q}$, $\,\bar{\omega}_{\mu \brp\brq}=(\bre^{-1})_{\brp}{}^{\nu}\trd_{\mu}\bre_{\nu \brq}$, \textit{\,etc.}  \\

After gauge fixing, $\bar{e}_{\mu}{}^{\brp}=e_{\mu}{}^{p}$, with the identification,  $\bar{\eta}_{\brp\brq}=-\eta_{pq}$, which breaks 
$\Ott$ to ${\mathbf{O}(10)\rtimes\GL(10)}$ \cite{Jeon:2011cn}, and the pair of local Lorentz symmetries to a single one, $\sot\times\osot\rightarrow\sot$,  it is straightforward to check  that 
the supersymmetric DFT Lagrangian (\ref{L024}) coincides with the $10D$ ${\cN=1}$ supergravity Lagrangian  (\ref{10DSUGRA}).  Further, the DFT supersymmetry~(\ref{SUSY}) agrees  with the $10D$ ${\cN=1}$ supersymmetry~(\ref{10DSUSY}),  up an $\sot$ local Lorentz transformation corresponding to the former of $\sot\times\osot$ having  the  parameter,  
\be 
-i\bar{\varepsilon}\gamma_{p}\psi_{q}+i\bar{\varepsilon}\gamma_{q}\psi_{p}\,.
\ee
~\\~\\

\subsection{Fierz identities}
Relevant  Fierz identities  include 
\be
\bar{\rho}\gamma^{pqr}\rho(\bar{\rho}\gamma_{pqr})_{\alpha}=0\,, 
\label{r4}
\ee
~\\
\be
\textstyle{\frac{1}{16}}\brrho{\gamma^{pqr}\rho}\bar{\psi}_{\brp}\gamma_{pqr}\psi^{\brp}=
\bar{\rho}{\gamma_{pq}\psi_{\brp}}\brrho\gamma^{pq}\psi^{\brp}\,,
\label{rrpp}
\ee
~\\
\be
\ba{l}
- \frac{1}{1728} \bar{\psi}^{m} \gamma^{n p q} \psi_{m} \bar{\rho} \gamma_{n p q} \rho - \frac{19}{864} \bar{\rho} \gamma^{m n p} \rho \bar{\psi}_{m} \gamma_{n} \psi_{p} + \frac{1}{12} \bar{\rho} \psi^{m} \bar{\rho} \gamma_{m n} \psi^{n}-\frac{11}{864} \bar{\rho} \gamma^{m n} \psi^{p} \bar{\rho} \gamma_{m n} \psi_{p} \\
- \frac{7}{216} \bar{\rho} \gamma^{m n} \psi^{p} \bar{\rho} \gamma_{n p} \psi_{m} - \frac{1}{432} \bar{\rho} \gamma^{m n} \psi^{p} \bar{\rho} \gamma_{m n p q} \psi^{q} - \frac{1}{864} \bar{\rho} \gamma^{m n p q} \psi_{q} \bar{\rho} \gamma_{m n p r} \psi^{r} \\
- \frac{1}{1728} \bar{\rho} \gamma^{m n p} \rho \bar{\psi}^{q} \gamma_{q r m n p} \psi^{r} - \frac{1}{144} \bar{\rho} \gamma_{mn}\psi^{n} \bar{\rho} \gamma^{m p} \psi_{p}
- \frac{7}{864} \bar{\rho} \gamma^{m n p} \rho \bar{\psi}_{n} \gamma_{m p q} \psi^{q} = 0\,,
\ea
\label{p2}
\ee
~\\
\be
\ba{l}
- \frac{1}{12} \bar{\rho} \psi^{m} \bar{\psi}_{m} \gamma^{n} \psi_{n} + \frac{13}{1728} \bar{\psi}^{m} \gamma^{n p q} \psi_{m} \bar{\rho} \gamma_{p q} \psi_{n} + \frac{19}{288} \bar{\psi}^{m} \gamma^{n} \psi^{p} \bar{\rho} \gamma_{m p} \psi_{n} - \frac{1}{16} \bar{\psi}^{m} \gamma^{n} \psi^{p} \bar{\rho} \gamma_{n p} \psi_{m} \\
- \frac{1}{72} \bar{\psi}^{m} \gamma_{m} \psi^{n} \bar{\rho} \gamma_{n p} \psi^{p} - \frac{1}{1728} \bar{\psi}^{m} \gamma^{n p q} \psi_{m} \bar{\rho} \gamma_{n p q r} \psi^{r} + \frac{25}{864} \bar{\psi}^{m} \gamma^{n} \psi^{p} \bar{\rho} \gamma_{m n p q} \psi^{q} \\
- \frac{5}{1728} \bar{\rho} \gamma^{m n} \psi^{p} \bar{\psi}^{q} \gamma_{q r p m n} \psi^{r} - \frac{1}{1728} \bar{\rho} \gamma^{m n p q} \psi_{q} \bar{\psi^{r}} \gamma_{r s m n p} \psi^{s} + \frac{13}{432} \bar{\rho} \gamma^{m n} \psi^{p} \bar{\psi}_{m} \gamma_{n p q} \psi^{q} \\
+ \frac{7}{864} \bar{\rho} \gamma^{m n} \psi^{p} \bar{\psi}_{p} \gamma_{m n q} \psi^{q} + \frac{1}{864} \bar{\rho} \gamma^{m n p q} \psi_{q} \bar{\psi}_{n} \gamma_{m p r} \psi^{r} = 0\,,
\ea
\label{p3}
\ee
and
\be
\ba{l}
- \frac{11}{216} \bar{\psi}^{m} \gamma^{n p q} \psi_{m} \bar{\psi}_{p} \gamma_{n} \psi_{q} + \frac{73}{432} \bar{\psi}^{m} \gamma^{n} \psi^{p} \bar{\psi}_{m} \gamma_{n} \psi_{p} - \frac{71}{432} \bar{\psi}^{m} \gamma^{n} \psi^{p} \bar{\psi}_{n} \gamma_{m} \psi_{p} - \frac{1}{144} \bar{\psi}^{m} \gamma_{n} \psi^{n} \bar{\psi}_{m} \gamma^{p} \psi_{p}\\
+ \frac{1}{108} \bar{\psi}^{m} \gamma^{n} \psi^{p} \bar{\psi}^{q} \gamma_{q r m n p} \psi^{r} - \frac{1}{108} \bar{\psi}^{m} \gamma^{n p q} \psi_{m} \bar{\psi}_{n} \gamma_{p q r} \psi^{r} - \frac{1}{12} \bar{\psi}^{m} \gamma^{n} \psi^{p} \bar{\psi}_{m} \gamma_{n p q} \psi^{q} \\
+ \frac{1}{144} \bar{\psi}^{m} \gamma^{n} \psi^{p} \bar{\psi}_{n} \gamma_{m p q} \psi^{q} + \frac{1}{864} \bar{\psi}_{m} \gamma^{m n p q r} \psi_{n} \bar{\psi}_{p} \gamma_{q r s} \psi^{s} - \frac{1}{432} \bar{\psi}^{m} \gamma^{n p q} \psi_{q} \bar{\psi}_{n} \gamma_{m p r} \psi^{r}\\
+ \frac{1}{216} \bar{\psi}^{m} \gamma^{n p q} \psi_{q} \bar{\psi}_{m} \gamma_{n p r} \psi^{r} - \frac{1}{192} \bar{\psi}^{q} \gamma^{m n p} \psi_{q} \bar{\psi}^{r} \gamma_{m n p} \psi_{r} =0\,.
\ea
\label{p4}
\ee
\end{widetext}

\begin{thebibliography}{99}
\bibitem{Buscher:1985kb} 
T.~H.~Buscher,
Phys. Lett. B {\bf 159} (1985) 127.
  
\bibitem{Buscher:1987sk}
T.~H.~Buscher,
  Phys.\ Lett.\  B {\bf 194} (1987) 59.
  
  
\bibitem{Buscher:1987qj}
T.~H.~Buscher,
Phys.\ Lett.\ B {\bf 201} (1988) 466.
  
\bibitem{Giveon:1988tt}
  A.~Giveon, E.~Rabinovici and G.~Veneziano,
  Nucl.\ Phys.\  B {\bf 322} (1989) 167.
  
  

\bibitem{Tseytlin:1990nb}
  A.~A.~Tseytlin,
  Phys.\ Lett.\  B {\bf 242}, 163 (1990).

\bibitem{Tseytlin:1990va}
  A.~A.~Tseytlin,
  Nucl.\ Phys.\  B {\bf 350}, 395 (1991).



\bibitem{Siegel:1993xq}
  W.~Siegel,
  Phys.\ Rev.\  D {\bf 47}, 5453 (1993).

\bibitem{Siegel:1993th}
  W.~Siegel,
  Phys.\ Rev.\  D {\bf 48}, 2826 (1993).
  
  
\bibitem{Hull:2009mi}
  C.~Hull and B.~Zwiebach,
  JHEP {\bf 0909}, 099 (2009).

\bibitem{Hull:2009zb}
  C.~Hull and B.~Zwiebach,
  JHEP {\bf 0909}, 090 (2009).

\bibitem{Hohm:2010jy}
  O.~Hohm, C.~Hull and B.~Zwiebach,
  JHEP {\bf 1007}, 016 (2010).
  [arXiv:1003.5027 [hep-th]].



\bibitem{Hohm:2010pp}
  O.~Hohm, C.~Hull and B.~Zwiebach,
  JHEP {\bf 1008}, 008 (2010).
  [arXiv:1006.4823 [hep-th]].


  
  
  
  
  
\bibitem{Kwak:2010ew}
  S.~K.~Kwak,
  JHEP {\bf 1010} (2010) 047.


\bibitem{Jeon:2010rw}
  I.~Jeon, K.~Lee and J.-H.~Park,
  JHEP {\bf 1104} (2011) 014.
  [arXiv:1011.1324 [hep-th]].
  


\bibitem{Jeon:2011cn}
  I.~Jeon, K.~Lee, J.-H.~Park,
  Phys.\ Rev.\  {\bf D84 } (2011)  044022.
  [arXiv:1105.6294 [hep-th]].


\bibitem{Jeon:2011vx}
  I.~Jeon, K.~Lee and J.-H.~Park,
 JHEP {\bf 11}   (2011)   025. 
 [arXiv:1109.2035 [hep-th]].



\bibitem{Jeon:2011kp}
  I.~Jeon, K.~Lee, J.~-H.~Park,
  Phys.\ Lett.\  {\bf B701 } (2011)  260-264.
  [arXiv:1102.0419 [hep-th]].





\bibitem{Hohm:2010xe}
  O.~Hohm, S.~K.~Kwak,
  J.\ Phys.\ A {\bf A44 } (2011)  085404.


\bibitem{Hohm:2011ex}
  O.~Hohm, S.~K.~Kwak,
  JHEP {\bf 1106 } (2011)  096.

\bibitem{Berman:2011pe}
  D.~S.~Berman, H.~Godazgar, M.~J.~Perry,
  Phys.\ Lett.\  {\bf B700 } (2011)  65-67.
 [arXiv:1103.5733 [hep-th]].

  


\bibitem{Thompson:2011uw}
  D.~C.~Thompson,
  JHEP {\bf 1108 } (2011)  125.

\bibitem{Copland:2011yh}
  N.~B.~Copland,
  Nucl.\ Phys.\  {\bf B854 } (2012)  575-591.

\bibitem{Hohm:2011zr}
  O.~Hohm, S.~K.~Kwak, B.~Zwiebach,
  Phys.\ Rev.\ Lett.\  {\bf 107 } (2011)  171603.
  [arXiv:1106.5452 [hep-th]].

\bibitem{Hohm:2011dv}
  O.~Hohm, S.~K.~Kwak, B.~Zwiebach,
  JHEP {\bf 1109 } (2011)  013.
  [arXiv:1107.0008 [hep-th]].

\bibitem{Albertsson:2011ux}
  C.~Albertsson, S.~-H.~Dai, P.~-W.~Kao, F.~-L.~Lin,
  JHEP {\bf 1109 } (2011)  025.
  [arXiv:1107.0876 [hep-th]].

\bibitem{Hohm:2011cp}
  O.~Hohm, S.~K.~Kwak,
  [arXiv:1108.4937 [hep-th]].

\bibitem{Kan:2011vg}
  N.~Kan, K.~Kobayashi, K.~Shiraishi,
    [arXiv:1108.5795 [hep-th]].

\bibitem{Aldazabal:2011nj}
  G.~Aldazabal, W.~Baron, D.~Marques, C.~Nunez,
   [arXiv:1109.0290 [hep-th]].


\bibitem{Geissbuhler:2011mx}
  D.~Geissbuhler,
   [arXiv:1109.4280 [hep-th]].

\bibitem{Berman:2011kg}
  D.~S.~Berman, E.~T.~Musaev, M.~J.~Perry,
  [arXiv:1110.3097 [hep-th]].


\bibitem{Copland:2011wx}
  N.~B.~Copland,
  [arXiv:1111.1828 [hep-th]].







\bibitem{Courant}
T. Courant, ``Dirac Manifolds,"  Trans. Amer. Math. Soc. {\bf 319:} 631-661, 1990.



\bibitem{Gualtieri:2003dx}
  M.~Gualtieri, Ph.D. Thesis,   Oxford University, 2003.
  arXiv:math/0401221.

  
\bibitem{Grana:2008yw}
M.~Grana, R.~Minasian, M.~Petrini and D.~Waldram,
JHEP {\bf 0904} (2009) 075.









 
     




  

\bibitem{Hitchin:2004ut}
  N.~Hitchin,
  Quart.\ J.\ Math.\ Oxford Ser.\  {\bf 54}, 281 (2003).

\bibitem{Hitchin:2010qz}
  N.~Hitchin,
  arXiv:1008.0973 [math.DG].
  
  

   
   
  
\bibitem{Coimbra:2011nw}
  A.~Coimbra, C.~Strickland-Constable, D.~Waldram,
    [arXiv:1107.1733 [hep-th]].


\bibitem{Berman:2010is}
  D.~S.~Berman and M.~J.~Perry,
 [arXiv:1008.1763 [hep-th]].

\bibitem{Berman:2011cg}
  D.~S.~Berman, H.~Godazgar, M.~Godazgar, M.~J.~Perry,
   [arXiv:1110.3930 [hep-th]].




\bibitem{Henneaux:1992ig}
  M.~Henneaux, C.~Teitelboim,
  Princeton, USA: Univ. Pr. (1992) 520 p.
  





\bibitem{LPTENS-78-10}
  E.~Cremmer, B.~Julia and J.~Scherk,
  Phys.\ Lett.\ B\ {\bf 76} (1978) 409.
  
  
    
\bibitem{Chamseddine:1980cp}
  A.~H.~Chamseddine,
  Nucl.\ Phys.\  B {\bf 185} (1981) 403.

 

      
\bibitem{Bergshoeff:1981um}
 E.~Bergshoeff, M.~de Roo, B.~de Wit and P.~van Nieuwenhuizen,
 Nucl.\ Phys.\  B {\bf 195} (1982) 97.
 
 
  
\bibitem{Hohm:2011nu}
  O.~Hohm and S.~K.~Kwak,
  arXiv:1111.7293 [hep-th].
    
    
    
    

\bibitem{VIDEO}
{ A  video presentation of  this work  is  available at}\\
{\scriptsize{\textsf{http://www.newton.ac.uk/programmes/BSM/seminars/021611301.html}}}

  
  
  \end{thebibliography}
\end{document}